\documentclass[twocolumn,pra,aps,showpacs]{revtex4b5}
\usepackage{epsfig}
\def\be{\begin{equation}}
\def\ee{\end{equation}}

\begin{document}
\title{Bloch Waves and Bloch Bands of Bose-Einstein Condensates 
in Optical Lattices}
\author{Biao Wu, Roberto B. Diener, and Qian Niu}
\affiliation{Department of Physics, The University of Texas at Austin,
Austin, Texas 78712-1081}
\date{\today}
\begin{abstract}
\vskip10pt
Bloch waves and Bloch band of Bose-Einstein Condensates 
in optical lattices are studied. We provide further evidence
for the loop structure in the Bloch band, and compute
the critical values of the mean-field interaction strength
for the Landau instability and the dynamical instability.
\end{abstract}

\maketitle

\section{Introduction}
Bose-Einstein condensates (BECs) in optical lattices have been
attracting increasing attention from both theorists 
\cite{nlz,stable,Smerzi,Carr,Bronski,Choi,Holthaus} and 
experimentalists \cite{Kasevich,Burger}. People are interested in how the 
interaction and coherence of this system affect the interesting phenomena 
observed with dilute cold atoms in optical lattices \cite{Bha},
such as Landau-Zener tunneling and Bloch oscillations. 
Recent studies have shown that these phenomena are indeed strongly
influenced by the interaction between atoms.
A series of novel effects have been discovered, 
including the nonlinear Landau-Zener tunneling \cite{nlz}, 
the breakdown of Bloch oscillations \cite{stable,Smerzi}, 
and dynamical instability \cite{stable,Carr}. 
There are similar nonlinear periodic systems in other fields, for example,
the system of the nonlinear guided waves in a periodic layered medium 
\cite{Kivshar}.

In a one dimensional optical lattice created by two 
counter-propagating off-resonance laser beams, 
a BEC is essentially a one dimensional system when the lateral motion can 
be either neglected \cite{Choi} or confined \cite{Burger}.
Its grand canonical Hamiltonian is 
\be\label{eq:h}
H=\!\int_{-\infty}^{\infty}\!{\rm d}x~\{\psi^*(-{1\over 2}
{\partial^2\over \partial x^2}+v\cos{x})\psi+{c\over 2}|\psi|^4-\mu|\psi|^2\},
\ee
where $\psi$ is the macroscopic wave function of the BEC. In
the above equation, all the variables are scaled to be dimensionless 
with the system's basic parameters, the atomic mass $m$,
the wave number $k_L$ of the two laser lights, and the average 
density $n_0$ of the BEC. The
strength of the periodic potential $v$ is in units of
${4\hbar^2 k_L^2\over m}$, the wave function $\psi$ in units of $\sqrt{n_0}$,
$x$ in units of ${1\over 2k_L}$, and $t$ in units of
${m\over 4\hbar k_L^2}$. The coupling constant $c={\pi n_0 a_s\over k_L^2}$,
where $a_s>0$ is the $s$-wave scattering length. 
A two dimensional version of this system has also received
some attention \cite{twod}.

In this Brief Report, we study the Bloch bands and Bloch waves
of a BEC in an optical lattice,  and present
new results that we were unable to obtain in 
our previous studies in Refs.\cite{nlz,stable}. 
These new results are possible now due primarily to a new development,
an exact solution found in Ref.\cite{Carr}.
As Bloch bands and Bloch waves are the two most important
concepts in understanding a linear periodic system, 
they shall also play crucial roles in the physics
of the nonlinear periodic system (\ref{eq:h}).
Bloch waves are the extremum states of
the Hamiltonian (\ref{eq:h}) of the form
\be\label{eq:bloch}
\psi(x,t)=e^{ikx}\phi_k(x)\,,
\ee
where $\phi_k(x)$ is a periodic function of period $2\pi$ and
$k$ is the Bloch wave number. Each Bloch wave state 
(\ref{eq:bloch}) satisfies the time-independent  Gross-Pitaevskii equation
\be\label{eq:nlsp}
-{1\over 2}({\partial\over \partial x}+ik)^2\phi_k
+c|\phi_k|^2\phi_k+v\cos x\,\phi_k=\mu\,\phi_k\,,
\ee
as can be verified by variation of the Hamiltonian (\ref{eq:h}).
Bloch bands are given by the set of eigenenergies $\mu(k)$. 
\begin{figure}[!htb]
\begin{center}
\resizebox *{6cm}{4.8cm}{\includegraphics*{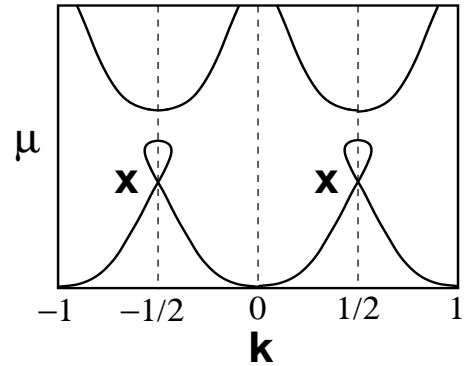}}
\end{center}
\caption{Schematic drawing of the first and second Bloch bands
of a BEC in an optical lattice when $c> v$.}
\label{fig:lobandk}
\end{figure}

\section{Bloch Bands}
In Ref.\cite{nlz}, we studied the tunneling between the 
two lowest bands to see how it is affected by the interaction.
We found that the tunneling is described by a revised 
Landau-Zener model, which we call the nonlinear Landau-Zener model. 
This model predicts a dramatic change in the band structure, which is a loop
appearing at the Brillouin zone edge $k=\pm 1/2$ 
for $c/v > 1$ (see Fig\,.\ref{fig:lobandk}).
A direct consequence of this loop structure is the breakdown
of the Bloch oscillations due to the non-zero adiabatic tunneling into
the upper band. 

The loop structure is confirmed by an exact solution found 
recently by Bronski {\it et al.}(Eq.(10) of Ref.\cite{Carr}),
which assumes a much 
simpler form in terms of our notations,
\be\label{eq:x}
\psi_B(x)=a_+e^{i{x\over 2}} + a_-e^{-i{x\over 2}}\,,
\ee
where
$a_\pm={\sqrt{c-v}\pm\sqrt{c+v}\over2\sqrt{c}}$. Substituting
it into Eq.(\ref{eq:nlsp}), we have $\mu={1\over 8}+c$.
This solution only exists when $c\ge v$, and is a Bloch wave at the
edge of the Brillouin zone, $k=1/2$.
This Bloch wave carries a non-zero velocity, ${\sqrt{c^2-v^2}\over 2c}$, 
while its complex conjugate has an opposite velocity. This is in sharp 
contrast with the behavior in a linear periodic system, 
in which Bloch waves at the zone edge always have zero velocity. 
This difference confirms the looped band structure. 
The solution $\psi_B$ and its complex conjugate
are the two degenerate states at the crossing point 
$X$ (Fig\,.\ref{fig:lobandk}). 
The non-zero velocity carried by this Bloch wave is a 
manifestation of superfluidity of BEC. For free particles, 
the flow $e^{ix/2}$ is stopped completely by Bragg scattering 
from the periodic potential; for the BEC, the flow can no longer be 
stopped when the superfluidity is strong, that is, $c>v$.

\begin{figure}[!htb]
\begin{center}
\resizebox *{7.0cm}{7.0cm}{\includegraphics*{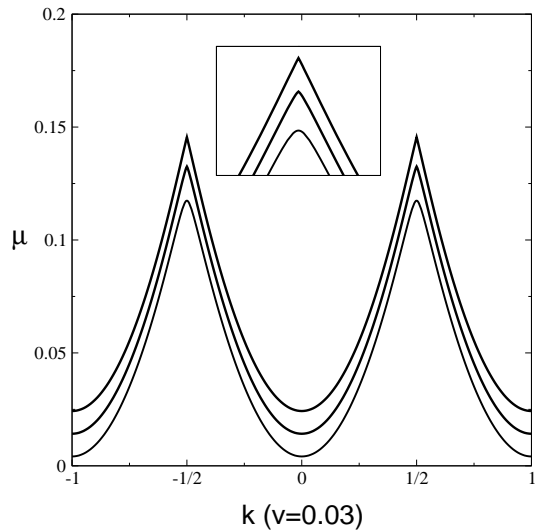}}
\end{center}
\caption{The lowest Bloch bands of 
of BECs in an optical lattice obtained by numerical calculation.
Top curve is for $c=0.05>v$; middle curve for $c=0.03=v$;
bottom curve for $c=0.01<v$. The inset is an enlarged version 
of the tips.}
\label{fig:gb}
\end{figure}

This loop structure is further supported by 
our numerical calculation of the lowest band $\mu(k)$,
as shown in Fig\,.\ref{fig:gb}. 
It is evident that the
slope ${\rm d}\mu/{\rm d}k$ at the zone edge $k=\pm 1/2$ 
becomes non-zero as the interaction strength
$c$ is increased over the periodic potential strength $v$,
a clear indication of the loop structure.
However, due to the limitation of our numerical method 
\cite{stable}, we are unable to produce directly the loop.
An improved numerical method is being developed to calculate
the loop and the higher Bloch bands.

\section{Stability of Bloch Waves}
In our second paper \cite{stable}, we studied the superfluidity
and stability of the Bloch waves in the lowest band (excluding
the loop). We found that the Bloch waves in the middle of the 
Brillouin zone represent super-flows,
and the other Bloch waves towards the zone edge have both a
Landau instability and a dynamical instability. Moreover, 
we found that these instabilities
can disappear from all these Bloch waves when the atomic
interaction is beyond certain critical values for a fixed 
lattice strength. For easy reference, we call the critical 
value for the Landau instability $c_L$, and the critical value
for the dynamical instability $c_d$. In that work,
we were unable to find these two critical values because
our numerical method was not good enough to find 
accurate Bloch waves at the zone edge. Now the exact
solution $\psi_B$ allows us to overcome the difficulty and
calculate these two critical values, $c_L$ and $c_d$. 
It is done by studying the stabilities of the Bloch wave $\psi_B$. 
Since the Bloch wave at the zone edge is the last one 
to become stable either in terms of the Landau instability 
or dynamically, the critical values of $c$ for $\psi_B$ to 
become stable are just $c_L$ and $c_d$. 

The physical significance of the two critical values, $c_L$ 
and $c_d$, lies in the way how the Bloch states at $k\neq 0$ are
achieved experimentally: the Bloch state at $k=0$ is
first prepared then driven to the desired Bloch states
at $k\neq 0$  by accelerating the optical
lattice\cite{Bha}. Therefore, as the only point
connecting the loop to the rest of the Bloch band,
a stable $\psi_B$ means that the Bloch states on the loop can be 
accessed and studied experimentally by accelerating the 
optical lattice.

\begin{figure}[!htb]
\begin{center}
\resizebox *{6cm}{6cm}{\includegraphics*{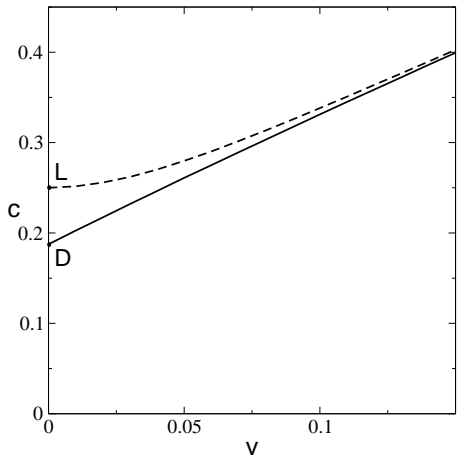}}
\end{center}
\caption{The critical values of $c$. The dashed line is $c_L$, 
the critical value of $c$ for all the Bloch waves in the lowest
band being super-flows; the solid line is $c_d$, above which
all the Bloch waves in the lowest band are dynamically stable.}
\label{fig:cv}
\end{figure}

We first study the Landau instability by analyzing how
the energy of the system deviates under a small perturbation. 
Since the system is periodic, we are allowed to
write the perturbation as
\be \label{eq:pert}
\psi=\psi_B+e^{i{x\over 2}}(u(x,q)e^{iqx}+v^*(x,q)e^{-iqx}),
\ee
where $q$ ranges between $-1/2$ and $1/2$, labelling the perturbation
mode, and the perturbation functions 
$u$ and $v$ have a periodicity of $2\pi$ in $x$.  
Then the energy deviation caused by this perturbation is
\be
\delta E=\int_{-\infty}^{\infty}{\rm d}x
\pmatrix{u^*,v^*}M(q)\pmatrix{u \cr v},
\ee
where
\be
M(q)=\pmatrix{{\mathcal L}(1/2+q) & c\phi_B^2 
\cr c\phi_B^{*2} & {\mathcal L}(-1/2+q)},
\ee
with
\be
{\mathcal L}(k)=-{1\over 2}({\partial \over \partial x} + ik)^2
-v\cos x+c-{1\over 8}
\ee
and
\be
\phi_B^2={c+\sqrt{c^2-v^2}\over 2c}-{v\over c}e^{-ix}+
{c-\sqrt{c^2-v^2}\over 2c}e^{-2ix}\,.
\ee
If $M(q)$ is positive definite for all $-1/2\le q\le 1/2$, the Bloch wave 
$\psi_B$ is a local minimum and a super-flow. Otherwise, $\delta E$ can be
negative for some $q$; the Bloch wave is a saddle point and
has a Landau instability. As already noticed in Ref.\cite{stable},
the positive definiteness of the matrices $M(q)$ for all
$q$'s is guaranteed by the positive definiteness of $M(0)$. 
Diagonalizing $M(0)$ for different values of $c$ with a fixed $v$, 
we obtain the critical value $c_L$, which is shown as a dashed line in 
Fig\,.\ref{fig:cv}. For the intersection point $L$ at $v=0$,
we have $c_L=1/4=(k=1/2)^2$. 

The dynamical stability of the Bloch wave $\psi_B$ is studied
by linearizing the Gross-Pitaevskii equation
\be\label{eq:nlsd}
i{\partial \psi\over \partial t}=
-{1\over 2}{\partial^2\psi\over \partial x^2}
+c|\psi|^2\psi+v\cos x\,\psi\,.
\ee
With a procedure similar to the  above, we arrive at
the linearized dynamical equation
\be\label{eq:first}
i{\partial\over \partial t}\pmatrix{u \cr v}=\sigma M(q)
\pmatrix{u \cr v},~~~\sigma=\pmatrix{I&0\cr0&-I}. 
\ee
The dynamical stability is determined by the matrix $\sigma M(q)$.
If all $\sigma M(q)$ for $-1/2\le q\le 1/2$ have no complex
eigenvalues, then $\psi_B$ is dynamically 
stable; otherwise, it is unstable. However, as pointed out
in Ref.\cite{stable}, the dynamical instability always
starts at the perturbation mode $q=1/2$. Therefore,
we only need to diagonalize $\sigma M(1/2)$ to find the critical value $c_d$.
The results are shown as the solid line in Fig\,.\ref{fig:cv}, where
the intersection $D$ at $v=0$ is precisely $c_d=3/16$.
This lower bound of the critical value $c_d$ simply means that
when $c<3/16$, any periodic potential brings the dynamical 
instability into the system. 

The value of $c_d$ at point $D$ is confirmed by analyzing
the limiting case $v\ll c$, where the matrix $\sigma M(1/2)$ can
be approximated with a $4\times 4$ matrix
\be
\sigma M(1/2)\approx \pmatrix{c-{1\over 8}&0&c&-v\cr
0&c+{3\over 8}&0&c\cr
-c&0&-{3\over 8}-c&0\cr
v& -c & 0 &{1\over 8}-c}\,.
\ee
The eigenvalues of this matrix can be found exactly; all of them
are real only when $c>3/16$. Note that the point $D$ must
be understood in a sense that $c_d\to 3/16$ as $v\to 0$ since
precisely at $v=0$ the system has no dynamical instability.
As one may get an impression from Fig.\ref{fig:cv} that the
asymptotic behavior of the two curves at large $v$ is linear,
we want to stress that it is not. Our numerical results
show that the asymptotic behavior undergoes very small oscillations
along a straight line, which we have no complete understanding.

Besides the existence of the loop structure shown
in Fig.\ref{fig:lobandk}, we have not discussed the properties
of the states on the loop due to the difficulty finding
these loop states accurately. Here we offer a glimpse of
these loop states and their properties. Around the crossing 
point $X$ (Fig.\ref{fig:lobandk}) $k=1/2+\epsilon$ 
($|\epsilon|\ll 1$), the Bloch wave can be approximated 
to the zeroth oder of $\epsilon$ as
\be
\psi(x)={\sqrt{h+1}+\sqrt{h-1}\over 2\sqrt{h}}e^{ikx}
-{\sqrt{h+1}-\sqrt{h-1}\over 2\sqrt{h}}e^{i(k-1)x}\,,
\ee
where $h={c\over v}+{1\over 2}\epsilon$. Numerical 
investigations show that most of these loop states
are all saddle points which can be either dynamically
stable or unstable. More numerical calculations
are needed to further confirm this.

Finally, we make two remarks. First, the way of defining 
Bloch waves and Bloch bands for the nonlinear system 
(\ref{eq:h}) at the beginning is a natural generalization
from the linear periodic system. Nevertheless, there is 
an essential difference due to the nonlinearity. 
In the linear system ($c=0$), the Bloch waves are the only
extremum states of its Hamiltonian or the only eigenfunctions
of Eq.(\ref{eq:nlsp}); for the nonlinear system
(\ref{eq:h}), there are possible extremum states that
are not Bloch waves.

Second, it is interesting to put the dynamical instability 
which is discussed in this report and in 
Ref.\cite{stable,Carr,Bronski} into perspective. 
Usually, a quantum dynamics is a regular motion because
it has discrete eigenvalues thus an almost periodic motion
no matter its corresponding classical dynamics is chaotic
or not. In this sense, quantum chaos has been called 
``pseudochaos''\cite{Casati}. On contrary, the 
dynamical instability that we have discussed is ``true''
quantum dynamical chaos that deserves more attention
in the future. 

On the other hand, with the Madelung transformation
$
\psi(x,t)=\rho(x,t) e^{iS(x,t)},
$
the nonlinear Schr\"odinger equation (\ref{eq:nlsd})
can be turned into a set of equations of fluid dynamics.
In this regard, the quantum dynamical instability should
be related to the turbulence in the fluid dynamics, 
and we may call it ``quantum turbulence''.

\vskip10pt
\acknowledgments{
This work is supported by the NSF, 
the Robert A. Welch Foundation, and the NSF of China.}

\end{document}